\documentclass[12pt]{amsart}
\usepackage{amsmath}
\usepackage{amsthm}
\usepackage{amsfonts}
\usepackage{graphicx}

\begin{document}

\title
{Dual variables for M-branes}
\author{JENS HOPPE}
\address{Braunschweig University, Germany}
\email{jens.r.hoppe@gmail.com}

\begin{abstract}
Motivated in parts by \cite{1},
relativistic extended objects will be described
by an (over-complete) set of generalized
coordinates and momenta that in some sense
are `dual' to each other.
\end{abstract}

\maketitle
%%%%%%%%%%%%%%%%%%%%%%%%%%%%%%%%%%%%%%%%%%%%%%%%%%%%%%%%%%%%%%%%%%%%%%%%%%%%%
%PAGE 1 PAGE 1 PAGE 1 PAGE 1 PAGE 1 PAGE 1 PAGE 1 PAGE 1 PAGE 1 PAGE 1 PAGE 1 
%%%%%%%%%%%%%%%%%%%%%%%%%%%%%%%%%%%%%%%%%%%%%%%%%%%%%%%%%%%%%%%%%%%%%%%%%%%%%
\noindent
$M$--dimensional extended objects in $D$--dimensional Minkowski--space
(see e.g. \cite{2} for a review) are describable by a Hamiltonian
whose potential is a homogenous polynomial of degree $2M$ in the
derivative of the coordinate fields. As always with nonlinear theories,
understanding their classical dynamics and possible quantization
rests on finding suitable variables. Hence, partly motivated by \cite{1}, 
let us consider
\begin{equation}\label{eq1} 
\begin{split}
L_{\beta} & = \int \big( p_{i/ \rho} \gamma^i_{\beta \alpha} + \dfrac{1}{M!}\lbrace x_{i_1}, \ldots, x_{i_M} \rbrace \gamma^{i_1 \ldots i_M}_{\beta \alpha} \big) \theta_{\alpha} \rho \, d^M \varphi \\
& =: \int \big( P^{\beta}_{\alpha} + Q^{\beta}_{\alpha} \big) \theta_{\alpha} \\
A & = - \dfrac{1}{(M-1)!} \dfrac{1}{2} \int \gamma^{i_2 \ldots i_M}_{\alpha \varepsilon} \theta_{\alpha} \lbrace \theta_{\varepsilon}, x_{i_2} \ldots x_{i_M} \rbrace \rho d^M \varphi  
\end{split}
\end{equation}
where the (canonically conjugate, classical) variables/fields, $x_{i=1 \ldots d}$ and $p_{j/ \rho}$, are functions on a compact $M$--dimensional manifold $\mathbb{M}$, the $\theta_{\alpha}(\varphi)$ non-dynamical `book--keeping' fermionic variables,  satisfying 
\begin{equation}\label{eq2} 
\theta_{\alpha}(\varphi)\theta_{\varepsilon}(\varphi') + \theta_{\varepsilon}(\varphi')\theta_{\alpha}(\varphi)
= \delta_{\alpha \varepsilon} \dfrac{\delta(\varphi, \varphi')}{\rho},
\end{equation}
\begin{equation}\label{eq3} 
\lbrace f_1 \ldots f_M \rbrace := \dfrac{1}{\rho} \varepsilon^{r_1 \ldots r_M} \partial_{r_1} f_1 \ldots \partial_{r_M} f_M
\end{equation}
(for any functions $f_r$ on $\mathbb{M}$; the density $\rho$, which according to \cite{M} can effectively be thought of as a constant, will for notational convenience be put $=1$), and the $\gamma^i$,
\begin{equation}\label{eq4} 
\gamma^i \gamma^j + \gamma^j \gamma^i = 2\delta^{ij} \mathbf{1}_{s \times s}
\end{equation}
assumed to be (constant) real symmetric $s \times s$ matrices, with $\gamma^{i_1 \ldots i_M}$ their standard fully antisymmetrized products ($= \gamma^{i_1} \ldots \gamma^{i_M}$, if all $i_r$ different).\\[0.15cm]
%%%%%%%%%%%%%%%%%%%%%%%%%%%%%%%%%%%%%%%%%%%%%%%%%%%%%%%%%%%%%%%%%%%%%%%%%%%%%
%PAGE 2 PAGE 2 PAGE 2 PAGE 2 PAGE 2 PAGE 2 PAGE 2 PAGE 2 PAGE 2 PAGE 2 PAGE 2 
%%%%%%%%%%%%%%%%%%%%%%%%%%%%%%%%%%%%%%%%%%%%%%%%%%%%%%%%%%%%%%%%%%%%%%%%%%%%%
Using (\ref{eq2}) one has
\begin{equation}\label{eq5} 
\begin{split}
L_{\beta}L_{\beta'} & + L_{\beta'}L_{\beta} \\ 
& = \int \big( p_i \gamma^i_{\beta \alpha} + \dfrac{1}{M!}\lbrace x_{i_1} \ldots x_{i_M} \rbrace \gamma^{i_1 \ldots i_M}_{\beta \alpha} \big)
       \big( p_k \gamma^k_{\beta' \alpha} + \dfrac{1}{M!}\lbrace x_{k_1} \ldots x_{k_M} \rbrace \gamma^{k_1 \ldots k_M}_{\beta' \alpha} \big)\\
& = \delta_{\beta \beta'} \int \vec{p}\,^2 + \int \dfrac{1}{M!}\lbrace x_{i_1} \ldots x_{i_M} \rbrace 
    \lbrace x_{k_1} \ldots x_{k_M} \rbrace \big(\dfrac{1}{M!} \gamma^{i_1 \ldots i_M}_{\beta \alpha} \gamma^{k_1 \ldots k_M}_{\beta' \alpha} + \beta \leftrightarrow \beta' \big)\\
& \quad + \int p_k \lbrace x_{i_1} \ldots x_{i_M} \rbrace \dfrac{1}{M!} \big( \gamma^{i_1 \ldots i_M}  \gamma^k \big)_{\beta \beta' + \beta' \beta} \\
& \cong \delta_{\beta \beta'}2\big(H = T+V \big)
\end{split}
\end{equation}
Due to the antisymmetrization $\lbrace x[_{i_1} \ldots x_{i_M} \rbrace \lbrace x_{k_1}] \ldots x_{k_M} \rbrace$ giving zero \cite{JH,ALMY} the second term on the rhs of (\ref{eq5}) gives
$\delta_{\beta \beta'}\dfrac{1}{M!}\int \sum\limits_{i_1\ldots i_M}\lbrace x_{i_1} \ldots x_{i_M} \rbrace^2$, as 
$\dfrac{1}{M!}\big( \gamma^{i_1} \ldots \gamma^{i_M} \gamma^{i_M}\ldots\gamma^{i_1} \big) = \mathbf{1}$, and the last term being zero on the reduced phase space of volume preserving (cp. \cite{JH}) functionals of $x$ and $p$ if $\gamma^{i_1 \ldots i_{M}k}$ is antisymmetric, i.e. if $\frac{M(M+1)}{2}$ is odd resp. $M = 1,2(\text{mod}\, 4)$.\\[0.15cm]
%%%%%%%%%%%%%%%%%%%%%%%%%%%%%%%%%%%%%%%%%%%%%%%%%%%%%%%%%%%%%%%%%%%%%%%%%%%%%
%PAGE 3 PAGE 3 PAGE 3 PAGE 3 PAGE 3 PAGE 3 PAGE 3 PAGE 3 PAGE 3 PAGE 3 PAGE 3 
%%%%%%%%%%%%%%%%%%%%%%%%%%%%%%%%%%%%%%%%%%%%%%%%%%%%%%%%%%%%%%%%%%%%%%%%%%%%%
On the other hand 
\begin{equation}\label{eq6} 
\begin{split}
L_{\beta}A & -AL_{\beta} \\
& = 
-\int \big( p_i \gamma^i_{\beta \alpha} + \dfrac{1}{M!}\lbrace x_{i_1} \ldots x_{i_M} \rbrace \gamma^{i_1 \ldots i_M}_{\beta \alpha} \big) \gamma^{k_2 \ldots k_M}_{\alpha \varepsilon} \dfrac{1}{(M-1)!}\lbrace \theta_{\varepsilon}, x_{k_2}\ldots x_{k_M}\rbrace\\
& = + \dfrac{1}{(M-1)!} \int \theta_{\varepsilon} \lbrace p_i, x_{k_2}\ldots x_{k_M} \rbrace \big(\gamma^i \gamma^{k_2 \ldots k_M} \big)_{\beta \varepsilon} \\
& \quad + \dfrac{1}{(M-1)!}\left\lbrace  \lbrace x_{i_1} \ldots x_{i_M} \rbrace, x_{k_1} \ldots x_{k_M}\right\rbrace 
\dfrac{1}{M!} \big( \gamma^{i_1 \ldots i_M} \gamma^{k_2 \ldots k_M} \big)_{\beta \varepsilon} \\
& \cong \int \big( \dot{Q}_{\beta \alpha} + \dot{P}_{\beta \alpha}\big)\theta_{\alpha}\\
& = \dot{L}_{\beta},
\end{split}
\end{equation}
where the first term in the middle of (\ref{eq6}) is zero on the reduced (volume--preserving) phase space whenever $i = k_r$ for some $r$ (in the first term) and in the second term due to the $\lbrace \, \rbrace$--identities (\cite{JH, ALMY}) unless $\lbrace k_2 \ldots k_M \rbrace \subset \lbrace i_1 \ldots i_M \rbrace$, (for which there are $M(M-1)\ldots 2 = M!$ possibilities).\\[0.15cm]
%%%%%%%%%%%%%%%%%%%%%%%%%%%%%%%%%%%%%%%%%%%%%%%%%%%%%%%%%%%%%%%%%%%%%%%%%%%%%
%PAGE 4 PAGE 4 PAGE 4 PAGE 4 PAGE 4 PAGE 4 PAGE 4 PAGE 4 PAGE 4 PAGE 4 PAGE 4 
%%%%%%%%%%%%%%%%%%%%%%%%%%%%%%%%%%%%%%%%%%%%%%%%%%%%%%%%%%%%%%%%%%%%%%%%%%%%%
Note that the equations of motion can be put into the form $\dot{Q} = \Omega P$, $\dot{P} = \Omega Q$,
\begin{equation}\label{eq7} 
\begin{split}
\dot{Q}^{\beta}_{\alpha} & = \pm \Omega^r_{\alpha \alpha'}\partial_r P^{\beta}_{\alpha'},\\
\dot{P}^{\beta}_{\alpha} & = \Omega^r_{\alpha \alpha'}\partial_r Q^{\beta}_{\alpha'},\\ 
\Omega^r_{\alpha \alpha'} & := \dfrac{1}{(M-1)!}\varepsilon^{r r_2 \ldots r_M}\partial_{r_2} x_{i_2}\ldots \partial_{r_M}x_{i_M} \gamma^{i_2 \ldots i_M}_{\alpha \alpha'}, 
\end{split}
\end{equation}
with the `matrix--vectorfield' $\Omega = (\Omega^r \partial_r) = (\Omega_{\alpha \alpha'})$ being divergence--free, and independent of $\beta$, and the $+$ sign applying to $M = (1) 2,5(\text{mod}\, 4)$, the $-$ sign to $M = 3,4(\text{mod}\, 4)$: 
\begin{equation}\label{eq8} 
\begin{split}
\Omega^r_{\alpha \alpha'} & \partial_r Q^{\beta}_{\alpha'}  \\
  & = \dfrac{1}{(M-1)!}\varepsilon^{r r_2 \ldots r_M}\partial_{r_2} x_{i_2}\ldots \partial_{r_M} x_{i_M} 
\gamma^{i_2 \ldots i_M}_{\alpha \alpha'}\dfrac{1}{M!}\partial_r \lbrace x_{k_1} \ldots x_{k_M} \rbrace \gamma^{k_1 \ldots k_M}_{\beta \alpha'}\\
 & =  \dfrac{1}{(M-1)!}\left\lbrace  \lbrace x_{k_1} \ldots x_{k_M} \rbrace, x_{i_2} \ldots x_{i_M}\right\rbrace 
\big( \dfrac{1}{M!} \gamma^{i_1 \ldots i_M} \gamma^{k_M \ldots k_1} \big)_{\alpha \beta}\\
 & =\dfrac{1}{(M-1)!}\left\lbrace  \lbrace x_{k_1} \ldots x_{k_M} \rbrace, x_{k_2} \ldots x_{k_M}\right\rbrace \gamma^{k_1}_{\alpha \beta}\\
& = \dot{p}_k \gamma^{k}_{\beta \alpha} \\
& = \dot{P}^{\beta}_{\alpha},
\end{split}
\end{equation}
and
\begin{equation}\label{eq9} 
\begin{split}
\dot{Q}^{\beta}_{\alpha} & = \frac{1}{(M-1)!} \lbrace p_i, x_{i_2} \ldots x_{i_M} \rbrace 
\gamma^{i i_2 \ldots i_M}_{\beta \alpha}\\
\Omega^r_{\alpha \alpha'}\partial_r P^{\beta}_{\alpha'} & =
\partial_r p_i \gamma^i_{\beta \alpha'}\frac{1}{(M-1)!} \varepsilon^{r r_2 \ldots r_M} \partial_{r_2}x_{i_2}\ldots \partial_{r_M}x_{i_M} \gamma^{i_2 \ldots i_M}_{\alpha \alpha'}\\
& = (-)^{\frac{(M-1)(M-2)}{2}} \lbrace p_i, x_{i_2} \ldots x_{i_M} \rbrace \big( \gamma^{i i_2 \ldots i_M} \big)_{\beta \alpha}.
\end{split}
\end{equation}
%%%%%%%%%%%%%%%%%%%%%%%%%%%%%%%%%%%%%%%%%%%%%%%%%%%%%%%%%%%%%%%%%%%%%%%%%%%%%
%PAGE 5 PAGE 5 PAGE 5 PAGE 5 PAGE 5 PAGE 5 PAGE 5 PAGE 5 PAGE 5 PAGE 5 PAGE 5 
%%%%%%%%%%%%%%%%%%%%%%%%%%%%%%%%%%%%%%%%%%%%%%%%%%%%%%%%%%%%%%%%%%%%%%%%%%%%%
It is then natural to look for `self--dual' solutions of the form
\begin{equation}\label{eq10} 
P^{\beta}_{\alpha} = A_{\alpha \alpha'} Q^{\beta}_{\alpha'}
\end{equation}
with (if $A$ is constant)
\begin{equation}\label{eq11} 
A \gamma^{i_2 \ldots i_M} A = (-)^{\frac{(M-1)(M-2)}{2}} \gamma^{i_2 \ldots i_M};
\end{equation}
choosing $A = \gamma := \gamma^1 \gamma^2 \ldots \gamma^d$, $\gamma^2 = (-)^{d\frac{(d-1)}{2}}$, 
$\gamma^T = (-)^{d\frac{(d-1)}{2}} \gamma$, $\gamma\gamma^i = (-)^{(d-1)}\gamma^i \gamma$ 
(hence $\gamma\gamma^{i_2 \ldots i_M}\gamma = (-)^{(M-1)(d-1)+d\frac{(d-1)}{2}}\gamma^{i_2 \ldots i_M}$)
one gets a condition on pairs $(M, d)$ which for $M=2$ is satisfied by $d= 2,5,6,9,10,\ldots$ while for $d = 3,6,7,10,11, \ldots$  if $M=3$. There is however no guarantee that (\ref{eq10}) is consistent for these allowed pairs $(M,d)$; e.g., for $d=2,\,M=2$, $\gamma^1 = \sigma_3$, $\gamma^2 = \sigma_1$, 
$\big(P^{\beta}_{\alpha}\big) = \big( p_i \gamma^i_{\beta \alpha} \big)=$
$\big(
\begin{smallmatrix}
p_1 & p_2 \\ p_2 & -p_1
\end{smallmatrix}
\big)
$,  
$\big(Q^{\beta}_{\alpha}\big) = \big( \lbrace x_1, x_2 \rbrace  \gamma^{12}_{\beta \alpha} \big)=$
$\big(
\begin{smallmatrix}
0 & x_{12} \\ -x_{12} & 0
\end{smallmatrix}
\big)
$,
$P^1_{\alpha} = \varepsilon_{\alpha'\alpha} Q^1_{\alpha'}$ gives $p_1 = x_{12}$, $p_2 = -Q^1_1 = 0$,
while $P^2_{\alpha} = \varepsilon_{\alpha \alpha'} Q^2_{\alpha'}$ implies $p_2 = Q^2_2 = 0$, 
$-p_1 = -Q^2_1 = x_{12}$.
While, related to quaternions and octonions, several special time--dependent solutions (see e.g. \cite{FL} and references therein) are known for the opposite (Euclidean) sign, which allows $d = 3,4,7,8, \ldots$ for $M=2$ and $d= 4,5,8,9$ for $M=3$
%%%%%%%%%%%%%%%%%%%%%%%%%%%%%%%%%%%%%%%%%%%%%%%%%%%%%%%%%%%%%%%%%%%%%%%%%%%%%
%PAGE 6 PAGE 6 PAGE 6 PAGE 6 PAGE 6 PAGE 6 PAGE 6 PAGE 6 PAGE 6 PAGE 6 PAGE 6 
%%%%%%%%%%%%%%%%%%%%%%%%%%%%%%%%%%%%%%%%%%%%%%%%%%%%%%%%%%%%%%%%%%%%%%%%%%%%%
it is important to notice the \textit{generally} significant structure of the equations of motion (\ref{eq7}), when written in the over--complete set of variables $Q$ and $P$ (with respect to which the Hamiltonian is quadratic, i.e. -was it not for the nontrivial symplectic form reflected by $\Omega-$ a set of harmonic oscillators).\\[0.15cm]
Finally, let me note that (\ref{eq11}), and the few lines following it,
has to hold only modulo terms that vanish when inserted
in the context of (\ref{eq7}); even in the case of the few already
known self--dual solutions (with the `wrong/euclidean' sign), which are not necessarily of the form (\ref{eq10}), as
generally mixing the $\beta$--indices, $P^{\beta}_{\alpha} = M_{\beta\alpha, \beta'\alpha'}Q^{\beta'}_{\alpha}$ the corresponding  consistency equation $M\Omega^r M = \pm \Omega^r$ has to (and will) hold only up to terms vanishing when acting on $\partial_r Q$.
%%%%%%%%%%%%%%%%%%%%%%%%%%%%%%%%%%%%%%%%%%%%%%%%%%%%%%%%%%%%%%%%%%%%%%%%%%%%%
%PAGE 7 PAGE 7 PAGE 7 PAGE 7 PAGE 7 PAGE 7 PAGE 7 PAGE 7 PAGE 7 PAGE 7 PAGE 7 
%%%%%%%%%%%%%%%%%%%%%%%%%%%%%%%%%%%%%%%%%%%%%%%%%%%%%%%%%%%%%%%%%%%%%%%%%%%%%
That the over--complete variables $Q$ and $P$ have relevant orthogonal subspaces plays a role already for finite degrees of freedom; comparing the time--evolution of $V^{\beta}_{\alpha a} = P^{\beta}_{\alpha a} + Q^{\beta}_{\alpha a}$ (see eq. (2), (6) of \cite{1}) with calculating the classical dynamical Poisson--brackets,
\begin{equation}\label{eq12} 
\begin{split}
\lbrace V^{\beta}_{\alpha a}, V^{\beta'}_{\alpha' a'} \rbrace & =
f_{aba'}x_{sb} \big( \gamma^{st}_{\beta \alpha} \gamma^{t}_{\beta' \alpha'} + \gamma^{st}_{\beta' \alpha'}\gamma^{t}_{\beta \alpha} \big) \\
& =: z^{\beta \beta'}_{\alpha a, \alpha' a'},
\end{split}
\end{equation}
hence
\begin{equation*}
\begin{split}
\dot{V}^{\beta}_{\alpha a} &= \lbrace V^{\beta}_{\alpha a},\, H  = \frac{1}{2}V^{\beta}_{\alpha' a'}V^{\beta}_{\alpha' a'} - \gamma^{t}_{\beta \beta} x_{t e} J_{e} \rbrace \\
& = z^{\beta \beta}_{\alpha a,\alpha' a'}V^{\beta}_{\alpha' a'} - \gamma^{t}_{\beta \beta} x_{t e}f_{aeb} V^{\beta}_{\alpha b}
\end{split}
\end{equation*}
implies e.g. that (for the fuzzy membrane in $d+2$ space--time dimensions \cite{7}, resp. $d+1$ Yang-Mills systems, e.g. \cite{8} for $d=3 \, , N=2$)
$$
f_{aba'}x_{sb} \big( \gamma^{st}_{\beta \alpha} \gamma^{t}_{\beta \alpha'} + \gamma^{st}_{\beta \alpha'} \gamma^{t}_{\beta \alpha} + \gamma^{s}_{\alpha \alpha'} -\gamma^{s}_{\beta \beta}\delta_{\alpha' \alpha} \big) V^{\beta}_{\alpha' a'} = 0.
$$

\vspace{1cm}
\noindent
\textbf{Acknowledgement.} 
I would like to thank L.Mason  for 
a conversation on twistors and self--duality,
M.Kontsevich for having pointed out calibrated geometry,
and S.Theisen for a discussion related to checking 
the linear relation between $P$ and $Q$ in an octonionic example.

\end{document}